\input epsfig.sty
\documentstyle[12pt]{article}

\def\PRL#1{{\it Phys.~Rev.~Lett.~}{\bf #1}}
\def\PRD#1{{\it Phys.~Rev.~}{\bf D#1}}

\def\NPB#1{{\it Nucl.~Phys.~}{\bf B#1}}

\def\PLB#1{{\it Phys.~Lett.~}{\bf B#1}}
\def\ARNPS#1{{\it Ann.~Rev.~Nucl.~Part.~Sci.~}{\bf #1}}

\def\RevModP#1{{\it Rev. Mod. Phys.}{\bf #1}}

\def\beq{\begin{equation}}
\def\eeq{\end{equation}}
\def\beqa{\begin{eqnarray}}
\def\eeqa{\end{eqnarray}}
\def\sm{Standard Model}
\def\tr{\tilde{t}_R}
\def\chr{\tilde{c}_R}
\def\tl{\tilde{t}_L}

\def\ur{\tilde{u}_R}

\def\egam{e^{-i\gamma}}

\def\ie{{\it i.e.} }
\def\Slash{\hskip -.6em/}

\def\gsim{{~\raise.15em\hbox{$>$}\kern-.85em
          \lower.35em\hbox{$\sim$}~}}
\def\lsim{{~\raise.15em\hbox{$<$}\kern-.85em
          \lower.35em\hbox{$\sim$}~}}

\def\etal{{\it et al.}}

\def\Slash#1{%
   \setbox0=\hbox{$#1$}\dimen0=\wd0\setbox1=\hbox{/}\dimen1=\wd1%
   \ifdim\dimen0>\dimen1\rlap{\hbox to \dimen0{\hfil/\hfil}}#1%
   \else\rlap{\hbox to \dimen1{\hfil$#1$\hfil}}/\fi}%

\begin{document}

\rightline{SLAC-PUB-7417}
\rightline{hep-ph/9702423}
\medskip
\rightline{February 1997}
\bigskip
\bigskip
\renewcommand{\thefootnote}{\fnsymbol{footnote}}

{\centerline{\bf SUPERSYMMETRIC BARYOGENESIS AT THE }}
{\centerline{{\bf ELECTROWEAK PHASE TRANSITION}
\footnotetext{Research supported
by the Department of Energy under contract DE-AC03-76SF00515}}}

\bigskip
{\centerline{Mihir P. Worah}}
\smallskip
\centerline {\it Stanford Linear Accelerator Center}
\centerline {\it Stanford University, Stanford, CA 94309}

\bigskip

{\centerline{\bf Abstract}}

We study 
the possibility of baryogenesis in the case of
supersymmetry breaking with large mixing between the $\chr$ and $\tr$ 
or $\ur$ and $\tr$ squarks resulting in one
light right-handed up-type squark mass eigenstate. We argue that 
in this case the
electroweak phase transition will be first order, 
and that large phases already present in the quark mass matrices  
can generate a baryon asymmetry of the correct
magnitude without introducing any new phases specifically for this
purpose. 
We study in detail a particular ansatz for supersymmetry breaking and
CP violation where there is only one CP violating phase in the theory:
in the up-type quark mass matrix. We study the constraints placed on
this model by baryogenesis and flavor physics.
This scenario has robust implications for low energy
flavor phsyics including 
$D^0-\bar D^0$ mixing and an electric dipole moment for the neutron
that are close to the experimental bounds, and CP violation in 
the $B-\bar B$ system that is different from that in the
Standard Model. 

\newpage

\section{Introduction}
\renewcommand{\thefootnote}{\arabic{footnote}}
\setcounter{footnote}{0}

The three physical requirements needed for baryogenesis
\cite{sakharov}: baryon number violation, CP violation, and a
departure from thermal equilibrium
all exist at the electroweak phase transition (EWPT).
The existence of baryon number
violating standard model field configurations at high temperatures
was demonstrated in \cite{KRS}.
CP violating interactions are known to exist in nature, and
a sufficiently
first order phase transition would provide the departure from
equilibrium needed in order to generate the baryon asymmetry and then
shut off the baryon number violation fast enought to avoid washing 
it out. The possibility of baryogenesis occurring at the electroweak
phase transition is exciting because the physics occurs at $\simeq
100$ GeV, and is plausibly testable in terrestrial experiments.

An interesting attempt to generate the asymmetry using only the
fields and CP violation in the Standard Model was made in
\cite{Farrar-Shaposhnikov}, but it was later shown in
\cite{Huet-Sather} that decoherence effects 
prevent the generation of a sufficiently
large asymmetry in this case.
In addition, the phase
transition in the \sm\ isn't quite strong enough to preserve any
asymmetry that might have been generated. 
Thus it seems that one needs to go beyond the \sm\
in order to generate the baryon number asymmetry. Although many models
have been proposed that successfully accomplish this
\cite{ewbaryo}, there is no one commonly accepted or
most plausible paradigm. 

A well motivated extension of the \sm\
is the Minimal Supersymmetric Standard Model (MSSM).
Introducing the extra sparticle content of the MSSM
brings with it a plethora of new masses, mixing angles 
and CP violating phases 
associated with the soft breaking of supersymmetry. These are
constrained by the assumptions of no new CP
violation beyond that of the Cabibbo Kobayashi Maskawa (CKM) matrix,
universal supersymmetry breaking
squark and gaugino masses, and universal and proportional
trilinear scalar couplings. The phenomenology of this
constrained model has been extensively studied
\cite{const-susy}. While rich in implications for high energy collider
experiments, the constrained MSSM by construction has negligible 
effects on low energy flavor physics and baryogenesis. 

In a recent series of papers \cite{susy-ewpt1,susy-ewpt2} 
it has been shown that
if the assumption of universal squark masses was relaxed to allow a
sufficiently light $stop$, the strength of the electroweak
phase transition would be enhanced over that in the \sm, allowing
any baryon asymmetry that might have been generated at the phase
transition to persist until today. 
This light $stop$ would have to be predominantly the $\tr$ because a
light $\tl$ would require a degenerate $\tilde{b}_L$ to avoid
conflicts with electroweak observables like the $\rho$ parameter, 
and generating a light $stop$ due
to large $\tl - \tr$ mixing weakens the strength of the EWPT.
One could then introduce a small phase
in the left-right $stop$ mixing term ($A_t$), which would
provide the CP violation required to generate a baryon
asymmetry \cite{huet-nelson} 
(models that have a large phase in the
diagonal trilinear term $A_t$ have the possibility of
generating too large a nucleon EDM due to an RGE induced phase in 
$A_u$ \cite{garristo-wells}).
Possible aesthetic objections to this
scenario are the presence of an extremely small or negative 
supersymmetry breaking mass squared for $\tr$, 
and the introduction of a new small CP violating phase. 

In this paper we suggest a model that generates a baryon asymmetry by
a generalization of the mechanism first proposed in
\cite{huet-nelson}, which does not suffer from the two unnatural
features mentioned above. 
The limits on neutral meson mixing constrain the mixing 
between the first two squark generations of either charge,
but do not limit the mixing between the first
and third or second and third generation 
up-type squarks \cite{luca}. 
Thus, there could be large $\chr-\tr$ or $\ur-\tr$ mixing   
in the soft susy breaking terms with universal diagonal terms \cite{nir2}. 
In such a scenario
one of the mass eigenstates is light, while still having a large
Yukawa coupling (proportional to the top quark mass) to the Higgs
boson, thus enhancing the electroweak phase transition as in
\cite{susy-ewpt1}. In addition, 
due to this texture of the $(RR)$ 
up squark mass matrix, the supersymmetry breaking 
$(LR)$ piece of
the squark mass matrix ($A$ term) is in general 
no longer proportional to the 
quark mass matrix, while the supersymmetry conserving part 
($\mu$ term) is \cite{hor-sym, dim-pom}. 
As a result of this non-proportionality
between the $A$ and $\mu$ terms the complex parameters
already present in the quark mass
matrix  give rise to a physical phase in the 
space and time dependent squark mass matrix
at the electroweak phase transition, which can 
generate a baryon number asymmetry of the correct magnitude.

The outline of our paper is as follows: 
in Sec. II we present the model.
Sec. III which is broken up into 4 sub-sections
discusses the details of the baryogenesis. Sec. IV studies the
implications for low energy flavor physics and Sec. V concludes.

\section{The Model} 

Consider the MSSM \cite{defMSSM}
with the most general soft supersymmetry breaking terms \cite{dim-pom}
\beq
{\cal L}_{MSSM}= {\cal L}_{SUSY}+{\cal L}_{soft}
\eeq
where
\beq
{\cal L}_{SUSY} = \int d^4\theta {\bf \Phi^{\dagger}}
                  e^{2gV}{\bf \Phi} + \int d^2\theta W({\bf
                  \Phi}) + h.c.
\label{lsusy}
\eeq
and
\beq
{\cal L}_{soft} = \int d^4\theta {\bf \Phi^{\dagger}}
                  [\bar\eta\Gamma^{*}+\eta\Gamma - \bar\eta\eta Z]
                  e^{2gV}{\bf \Phi}
             -\int d^4\theta{\bf\Phi^{T}}\frac{\Lambda}{2}{\bf\Phi} 
          + \int d^2\theta W'({\bf
                  \Phi}) + h.c.
\label{lsoft}
\eeq
In the  above equations 
${\bf\Phi}$ is a column vector of chiral superfields, $V$ are
the vector superfields, and $W({\bf\Phi})$ is the superpotential
containing bilinear and trilinear terms $M_{ij}$ and $Y_{ijk}$. In
Eq.~(\ref{lsoft}), $\eta$ is the spurion whose $vev = m_0\theta^2$ 
breaks
supersymmetry, $\Gamma$, $Z$ and $\Lambda$ are matrices of
supersymmetry breaking parameters, and $W'({\bf\Phi})$ is a
holomorphic function of the superfields with bilinear couplings
$M'_{ij}$ and trilinear couplings $Y'_{ijk}$.
This leads in component notation to 
\beq
-{\cal L}_{soft} = m^2_{ij}\phi^{*}_{i}\phi_{j} + 
                   \frac{1}{6}A_{ijk}\phi_i\phi_j\phi_k
                  +\frac{1}{2}B_{ij}\phi_i\phi_j 
	        %  + \frac{1}{2}\tilde M_i\tilde\lambda_i^2 
                  + h.c.
\eeq
where $\phi_i$ are the scalar fields and 
\beqa
A_{ijk} &=& m_0[Y'_{ijk}+Y_{ljk}\Gamma_{li}+Y_{ilk}\Gamma_{lj}+
                Y_{ijl}\Gamma_{lk}] \nonumber \\
B_{ij}  &=& m_0[M'_{ij}+M_{lj}\Gamma_{li}+M_{il}\Gamma_{lj}
                +m_0\Lambda_{ij}], \nonumber \\
m^2_{ij}&=& m_0^2[Z_{ij}+\Gamma^{*}_{il}\Gamma_{lj}],
%\tilde M_i &=& 
\label{components}
\eeqa
and we have ignored the gaugino mass terms.

Since the new
effects we consider are a result of the non-diagonal texture of the
right-handed up-type squark mass matrix squared 
($M^2_{\tilde u_{RR}}$) and the corresponding form of the
trilinear supersymmetry breaking term ($A_{\tilde u_{LR}}$) 
we concentrate primarily on these. Consider 
\beq
M^2_{\tilde u_{RR}} = m_0^2A_{U_{RR}}
\label{mrr}
\eeq
and 
\beq
A_{\tilde u_{LR}} = m_0 \lambda_U A_{U_{LR}}
\label{alr}
\eeq
where from Eq. (\ref{components}) $A_{U_{RR}}=
Z_{U_{RR}}+\Gamma^{\dagger}_{U_{RR}}\Gamma_{U_{RR}}$,
$\lambda_U$ is the matrix of 
Yukawa couplings for the up-type quarks, and
$A_{U_{LR}}=\lambda_U^{-1}Y_U' + \Gamma_{U_{RR}}$.
With the rest of the squark mass matrices diagonal and universal, 
the CP violating invariant responsible for
baryogenesis in this model is given by
\beq
J_{CP}=Im~Tr[A_{U_{LR}}^{\dagger}
             \lambda_U^{\dagger}\lambda_UA_{U_{RR}}]
\label{jcp}
\eeq
Since $\lambda_U^{\dagger}\lambda_U$ and $A_{U_{RR}}$ in  
Eq. (\ref{jcp}) are Hermitian, a necessary condition for baryogenesis
is that $A_{U_{LR}}$ not commute with either of these matrices.

In the constrained MSSM, 
$A_{U_{LR}}$ is proportional to the unit matrix at the
supersymmetry breaking scale, and is always Hermitian and commutes
with $\lambda_U^{\dagger}\lambda_U$ even after 
including RGE running to a lower scale.  
The scenario of supersymmetric baryogenesis with diagonal squark
masses \cite{huet-nelson} 
requires two small parameters beyond those of the constrained MSSM: 
a small supersymmetry breaking mass for $\tr$ $[A_{U_{RR}}(3,3)]$, and a
complex phase for the trilinear coupling 
$A_t$ [$A_{U_{LR}}(3,3)$].

Our approach is to notice that given the plausible assumption 
that at least part
of the CP violation seen in $K-\bar K$ mixing is due to \sm\ box
diagrams, $\lambda_U^{\dagger}\lambda_U$ in
Eq. (\ref{jcp}) already contains large phases and to try to use these
for baryogenesis without introducing 
CP violation into the supersymmetry breaking sector. All we need are
non-commuting  $A_{U_{LR}}$ and $A_{U_{RR}}$. 
This is possible, for example, in
models of horizontal symmetries due to the non-holomorphic terms
discussed in \cite{nir2,hor-sym}.  

We now study a specific realization of the scenario discussed above, 
making a series of assumptions in order to gain predictive power and
relate the CP violating phase responsible for baryogenesis to the one
observed in low energy meson decays. We
first assume that CP violation originates in the
supersymmetry conserving part of the lagrangian, ${\cal L}_{SUSY}$. In
the exact supersymmetric limit this corresponds to just one physical
phase in the mixing matrix for 3 families of quarks. We will limit
ourselves to this one phase, and make the following ansatz for the up
and down type
Yukawa coupling matrices:
\beq
\lambda_U = V^{\dagger} \hat \lambda_U V;~~~~
\lambda_D = \hat \lambda_D
\label{yukawas}
\eeq
where $V$ is the CKM matrix, and $\hat \lambda_U$ and $\hat\lambda_D$
are diagonal matrices. Consistent with our assumption of no CP
violation in the supersymmetry breaking sector, ${\cal L}_{soft}$,  
we set $W'({\bf\Phi})=0$%
{\footnote{In supergravity models with hidden sector
supersymmetry breaking one has $W'({\bf\Phi})=aW({\bf\Phi})$. Thus a
clear separation of the origin of CP violation demands
$W'({\bf\Phi})=0$.}}, 
and insist that the parameters $Z$, $\Gamma$,
and $\Lambda$ in Eq. (\ref{lsoft}) are real. The relative
phases between ${\cal L}_{SUSY}$ and ${\cal L}_{soft}$ will be
responsible for baryogenesis. While for the most general case there
are a large number of these relative phases that are physical, our
assumption makes them all proportional to the one fundamental 
phase in the quark mixing matrix. 

Notice, from
Eqs.~(\ref{components}) that it is the $\Gamma_{U_{RR}}$ part of  
$M^2_{\tilde u_{RR}}=m_0^2(Z_{U_{RR}}+\Gamma^{\dagger}_{U_{RR}}
\Gamma_{U_{RR}})$ that is responsible for $A_{\tilde u_{LR}}$ 
not being proportional to $\lambda_U$. 
Thus, in order to reduce the number of
arbitrary parameters we further set $Z_{U_{RR}} = 0$.
Using the ansatz
\beq
\Gamma_{U_{RR}}=
	                 \left(\begin{array}{ccc}
                     1& 0& 0\\
                     0& 1& x \\
                     0& y& 1 
                        \end{array} \right)
\label{texture}
\eeq
with $x,y \simeq 1$ in order to
have large $\chr - \tr$ mixing
\footnote{For the rest of the paper we will quote formulae relevant to
the case of $\chr-\tr$ mixing. The physics for the case of $\ur-\tr$
mixing is similar, and can be trivially generalized by moving the
parameters $x$ and $y$ from the (2,3) and (3,2) elements to the (1,3)
and (3,1) elements of $\Gamma_{U_{RR}}$.}
leads by Eq.~(\ref{components}) to 
\beq
M^2_{\tilde u_{RR}} = m_0^2A_{U_{RR}}=m_0^2
                     \left(\begin{array}{ccc}
                     1& 0& 0\\
                     0& 1+x^2& x+y \\
                     0& x+y& 1+y^2 
                        \end{array} \right)
\label{mrr2}
\eeq
and 
\beq
A_{\tilde u_{LR}} = m_0 \lambda_U A_{U_{LR}}=
                    m_0 \lambda_U\Gamma_{U_{RR}}
\label{alr2}
\eeq
We let the rest of the scalar mass squared matrices be universal, and
proportional to $m_0^2$ as in the constrained
MSSM. For the rest of the parameters that define the softly broken
MSSM we allow generic values
\beq 
m_0 \simeq |\mu| \simeq M_3 \simeq M_2 \simeq 100-300 ~GeV;~~~
\tan\beta\simeq 1
\label{params}
\eeq
Eqs.~(\ref{yukawas} - \ref{params}) along with universality for the
other scalar mass matrices define the model we will be studying.

We emphasize that we have made specific assumptions
about soft supersymmetry breaking and CP violation that allows us to
relate $A_{\tilde u_{LR}}$ to $\lambda_U$ and $M^2_{\tilde u_{RR}}$ 
($W', ~Z_{U_{RR}}=0$ and $\Gamma_{U_{RR}}$ is real).
It would be interesting if such a model of
supersymmetry breaking exists since, as we will show, 
it allows us to explain
both the CP violation seen in $K$ mesons and that required for
baryogenesis in terms of one complex parameter in the MSSM super
potential.
In order to do so we introduced two additional
parameters beyond those of the
constrained MSSM: the squark mixing parameters $x$ and $y$ which could
find their origin in models with horizontal symmetries.
Even if nature were such that some of the assumptions about
supersymmetry breaking and CP violation we have made did not hold, 
the rest of this paper serves
as a useful study of the possibility of electroweak baryogenesis for
the general case of non-universal and non-proportional 
supersymmetry breaking soft terms. In particular, 
all of our results would go through essentially unchanged if some
of the parameters that we set to zero actually turn out to be
non-vanishing but small.

\section{Baryogenesis}

First we will consider the strength of the EWPT and argue that a light
up-type squark produced due to $\chr - \tr$ mixing serves to enhance
the strength of the phase transition, 
possibly making it a first order one, and
allowing any baryon asymmetry produced during the phase transition to
persist to lower temperatures.

We then calculate the baryon asymmetry generated at the EWPT. 
We will not explicitly
state the approximations made in this approach, and refer the
reader to \cite{huet-nelson} for a discussion of the validity
of the assumptions and approximations made. 
The procedure is quite simple: 
first we calculate a CP violating source term for axial baryon 
($\chr + \tr$) number
induced by the passage of the bubble wall through
the plasma. Then we solve a set of coupled Boltzmann equations that
include the effects of diffusion, rapid particle number changing
interactions and the CP violating source. 
One finds then that there is an excess of $\chr + \tr$ number
generated in the symmetric phase. This biases the electroweak
sphalerons towards a production of net excess baryon number in the
symmetric phase, which is then stored in the broken phase as the
bubble sweeps through the plasma 
(the CP violating source term
acts as the charge potential of earlier models \cite{ckn1}, and the
effects of diffusion are important in moving the $\chr + \tr$
asymmetry out into the symmetric phase where sphalerons efficiently
convert this excess into a net excess baryon number \cite{ckn2}).

\subsection{The Electroweak Phase Transition}

The EWPT in the MSSM has been studied using two approaches: looking at
the daisy-improved effective potential \cite{early-ewpt,susy-ewpt1}
and
by lattice simulation of a dimensionally reduced effective theory
\cite{susy-ewpt2} with compatible results. The earlier studies
\cite{early-ewpt} working with the assumption of universal scalar
masses concluded that the situation in the MSSM was not much better
than that in the \sm~ where a strong enough first order phase
transition required an experimentally excluded Higgs boson mass.

Recently, however it has been shown \cite{susy-ewpt1,susy-ewpt2} that if
$\tan\beta$ is small, and the
right-handed stop, $\tr$ is light due to a sufficiently 
small supersymmetry breaking mass, the order of the EWPT can be
significantly enhanced over that in the \sm. 
The reason for this is quite simple. 
Scalars with vanishing or small Higgs independent mass 
contribute a cubic term to the scalar potential proportional
to their Higgs dependent mass, resulting in a finite temperature 
Higgs potential of the form
\beq
V(h,T) = -\frac{M^2(T)}{2}h^2 - \delta T h^3 + \frac{\lambda(T)}{4}h^4
\label{pot}
\eeq
with 
\beq
\delta = \frac{1}{12\pi}\sum_{i} g_i^3
\label{delta}
\eeq
where the sum is over all boson degrees of freedom
with zero Higgs independent mass,
and Higgs dependent mass defined by $m_i(h)=g_i h$.
The critical temperature for the phase transition, $T_0$
is defined as the temperature where
\beq
\frac{\partial^2 V}{\partial h^2}{\Bigg\vert}_{h=0}=0  
\Rightarrow M^2(T) = 0.
\label{to}
\eeq
This leads to a phase transition temperature $T_0 \sim 100$ GeV in
both the \sm, as well as the MSSM.
Minimizing the potential at the critical temperature,
\beq
\frac{\partial V}{\partial h}{\Bigg\vert}_{T=T_0}=0 
\Rightarrow -M^2(T_0)-3\delta T_0 h + \lambda(T_0)h^2=0
\label{minv}
\eeq
leads to the equation
\beq
\frac{h(T_0)}{T_0}=\frac{3\delta}{\lambda(T_0)}
\label{strength}
\eeq
for the field strength of the Higgs field $h$ at the
non-zero solution of Eq.~(\ref{minv}). 
Thus we see that it is the presence of the cubic term $\delta$
which forces the existence of two degenerate minima at the critical
temperature, and hence a first order phase transition.

Since the $\tr$ couples to the Higgs boson with
the large top quark Yukawa coupling, a light $\tr$ can significantly
enhance the strength of the electroweak phase transition.  
The quantity $h(T_0)/T_0$
is a measure of the strength of the phase transition.
The requirement on the strength of the phase transition
in order for the baryon asymmetry not to be erased by electroweak
sphalerons after the phase transition is \cite{fopt}
\beq
\frac{h(T_0)}{T_0} \geq 1
\label{strength2}
\eeq
To get a numerical estimate of the right-hand side of
Eq.~(\ref{strength}) one can use Eq.~(\ref{delta}) for $\delta$
with a vanishing supersymmetry breaking mass for $\tr$, and  
approximate $\lambda(T_0)$ by $\lambda(0) = m_h^2/2v^2$ where
$m_h$ is the physical Higgs boson mass, and $v = 246$ GeV is the zero
temperature vacuum expectation value (vev). This leads to the following
estimate for the strength of the EWPT in the \sm\ and the MSSM
for a Higgs boson mass of $75$ GeV
\beq
\left (\frac{h(T_0)}{T_0}\right )_{SM} \simeq 0.5;~~~~
\left (\frac{h(T_0)}{T_0}\right )_{MSSM} \simeq 3.
\eeq

Precisely this effect occurs if the lightest right-handed up-type
squark is an admixture of $\chr$ and $\tr$. The large mixing
generates a light eigenstate with small Higgs independent mass
%,\cf Eqs.~(\ref{mrr}, \ref{evals}),
without introducing any unnaturally small diagonal mass parameters. 
In addition, this lightest eigenstate will still have couplings to 
the Higgs boson proportional to the top quark Yukawa (modified by a
mixing angle of order 1), hence introducing a large cubic
term in the Higgs potential, and ensuring a first order phase
transition. We have confirmed these arguments for the model we propose
and find a strongly first order phase transition over large regions of
parameter space. 
For example using a Higgs mass of $75$ GeV, $m_0=200$ GeV and
$x=-1,~y=0$ in Eq.~(\ref{mrr}) leads to
\beq
\left (\frac{h(T_0)}{T_0}\right ) \simeq 1.5
\eeq
using the approximations mentioned above and with appropriate 
modifications to Eqs. (\ref{pot} - \ref{strength}) in order to account
for the non-zero supersymmetry breaking $stop$ mass.

We should mention two effects we have neglected in the arguments
of the previous two paragraphs that both tend to decrease the strength
of the phase transition.
The first is thermal masses for the squarks and
gauge bosons which act like non-zero soft susy breaking masses for 
the squarks and tend to screen the cubic term. This effect has been
included in both of \cite{susy-ewpt1}.
The second is including $L-R$ mixing terms between the squarks 
which generally 
reduces the field dependent mass of the lightest eigenvalue, hence
decreasing the magnitude of the cubic term. This effect was studied in
the first of \cite{susy-ewpt1} where they find a large enough
enhancement of the strength of the phase transition only if
$|m_0A_U+\mu/\tan\beta| \le m_0$. Given this condition, both find a
sufficiently first order phase transition for
$\tan\beta\simeq 1$ and a $stop$ mass $\le 175$ GeV. 
Although we have not studied the inclusion of these effects,
we will adopt these results and insist that $\mu$ is negative since
in our model both $A_U$ and $\tan\beta \simeq 1$, 
and use as our criterion for a phase transition
satisfying Eq. (\ref{strength2}) the condition $m_{\tilde u_1} \le 175$
GeV, where $m_{\tilde u_1}$ is the mass of the lightest up-type
squark.  

A final concern in models with small supersymmetry 
breaking squark masses is
the possibility of a negative mass squared for one of the mass
eigenstates resulting in a color and charge breaking vaccum. In order
to avoid this possibility we insist that the mass squared for the
lightest up-type squark be positive in the symmetric phase.

There has also been recent work suggesting that 2-loop QCD corrections
associated with $stops$ substantially enhances the strength of the
phase transition \cite{espinosa}. Such an effect would imply that our
constraints are conservative.

\subsection{The CP Violating Source Term}

In this sub-section we 
reproduce the salient points of the analysis
in \cite{huet-nelson}, which was limited to
looking at the third generation of squarks only, generalized to account
for $\chr-\tr$ mixing and non-proportional $A$ terms.

We define the CP violating source term for axial squark number
(actually quark + squark) as 
\beq
\gamma_{\tilde{Q}} = 
\frac{v_w}{\Delta}[J_{+}(\vec{x},t)+J_{-}(\vec{x},t)]^{0} 
\eeq
where $J_{\pm}^{\mu}$ are CP violating charge currents (in the plasma
frame) generated in a thickness $\Delta$ that the wall sweeps through
with velocity $v_w$. If we make an expansion for the currents in the
wall velocity $v_w$, and if we approximate the phase space
distribution for the squarks to be given by that in the symmetric
phase, we get 
\beq
\gamma_{\tilde{Q}}(z,t)=\gamma_{w} v_{w} N_c \frac{T^4}{4\pi^2} 
                   \times Im Tr[{\cal J}_{\tilde{Q}_L}{\cal I}_{L}
                                +{\cal J}_{\tilde{Q}_R}{\cal I}_{R}]
\eeq
the quantities ${\cal J}_{\tilde{Q}_L}{\cal I}_{L}$ and 
${\cal J}_{\tilde{Q}_R}{\cal I}_{R}$ can be related by a power series
expansion to $J_{CP}$ defined in Eq. (\ref{jcp}) (with a trivial
generalization to account for the left-handed squarks).
The coordinate $z=0$ defines the
edge of the symmetric phase, and $z=w$ defines the edge of the broken
phase, \ie $w$ is the thickness of the wall.
${\cal I}_{L,R}$ are $3 \times 3$ diagonal matrices with entries that
can be fit by the formula
\beq
[{\cal I}_{L,R}]_{ii}=\frac{1}{50}
\frac{T}{m_i}\frac{e^{m_i/T}}{(1-e^{m_i/T})^2}
\label{fit}
\eeq
where 
$m_i$ are the eigenvalues of $M^2_{\tilde{u}_{LL,RR}}$ in the
symmetric phase. Further,
\beq
{\cal J}_{\tilde{Q}_{L}}=-\frac{1}{z T^5}
                     U_{L}M^2_{\tilde{u}_{LR}}(z)
                     M^2_{\tilde{u}_{LR}}(0)^{\dagger}U_{L}^{\dagger};~
{\cal J}_{\tilde{Q}_{R}}=\frac{1}{z T^5}
                     U_{R}M^2_{\tilde{u}_{LR}}(z)^{\dagger}
                     M^2_{\tilde{u}_{LR}}(0)U_{R}^{\dagger}
\eeq
where 
\beq
M^2_{\tilde{u}_{LR}}(z) = m_0 \lambda_U\Gamma_{U_{RR}}v_2(z) + 
                          \mu \lambda_U v_1(z)
\label{mu-lr}
\eeq
and $U_{L,R}$ are the unitary matrices that diagonalize
$M^2_{\tilde{u}_{LL,RR}}$. Ignoring the $(LL)$ sector since it does
not contribute to the CP violating source term in our model, 
we have  
\beq
U_R = \left(\begin{array}{ccc}
1 & 0 & 0 \\
0 & c & s \\
0 & -s & c 
                        \end{array} \right)
\eeq 
with $c$
and $s$ parametrizing the sine and cosine of the $\chr-\tr$ mixing angle. 
Making an expansion in derivatives of the mass matrix (the thick
wall limit), Eq.~(\ref{mu-lr}) for 
$M^2_{\tilde{u}_{LR}}$, gives us
the following expression for the imaginary parts of the diagonal
elements in ${\cal J}_{\tilde{Q_R}}$:
\beqa
Im[{\cal J}_{\tilde{Q_R}}]_{ii} &=&\displaystyle{ 
                              \frac{4 m_0 \mu
                              M_{W}^2(z,T){\partial}_z\beta}    
                              {g^2 T^5}
                      Im[U_R\Gamma_{U_{RR}}^{\dagger}\lambda_{U}^{\dagger} 
                        \lambda_{U}U_R^{\dagger}]_{ii}} \nonumber \\
                         &\simeq&\displaystyle{
                              \frac{4 m_0 \mu
                              M_{W}^2(z,T)\Delta\beta}    
                              {w g^2 T^5}
                       Im[U_R\Gamma_{U_{RR}}^{\dagger}V^{\dagger}
                            \hat{\lambda}_{U}^{2}VU_R^{\dagger}]_{ii}}
\eeqa
where we have explicitly removed all the dependence on the Higgs vevs
to the pre-factor, so $\hat{\lambda}_{U}$ 
is the diagonal matrix of up-type quark
Yukawa couplings obtained from Eq.~(\ref{yukawas}), and we have replaced
$\partial_z\beta$ by $\Delta\beta /w$ where $w$ is the wall thickness,
and $\Delta\beta$ the variation of the Higgs vevs over the wall. 

For {\footnote{
Notice that this differs from the Wolfenstein parametrization
\cite{wolf} by the trivial replacement $\sigma=\sqrt{\rho^2+\eta^2}$ 
and by a rephasing of the third column. This then specifies the
flavor basis in which CP violation is defined as originating in the
supersymmetry conserving lagrangian. All low energy \sm\ calculations
would be invariant under the change from this basis to any other,
however in order for the full supersymmetric result to reamin
invariant, one would have to rephase also the squark mass matrices
appropriately.}}
\beq
V = \left(\begin{array}{ccc}
1-\lambda^2/2& \lambda & A\lambda^3\sigma \\
-\lambda& (1-\lambda^2/2) & A\lambda^2 e^{i\gamma} \\
A\lambda^3(1-\sigma \egam)& -A\lambda^2  & e^{i\gamma}
                        \end{array} \right)
\label{vkm}
\eeq 
We get 
\beq
Im[{\cal J}_{\tilde{Q}}]_{ii} = 
                              \frac{4 m_0 \mu
                              M_{W}^2(z,T)\Delta\beta}    
                              { w g^2 T^5}
\left(
\begin{array}{c}
0 \\
A\lambda^2 \lambda_t^2 \sin (\gamma)(x c^2 - y s^2) \\
-A\lambda^2 \lambda_t^2 \sin (\gamma)(y c^2 - x s^2) 
\end{array}
\right)
\label{cpsource}
\eeq
where $\lambda_t$ is the top quark Yukawa coupling, $x$ and $y$ are the
off-diagonal entries in the matrix $\Gamma_{U_{RR}}$, and
$\sin\gamma \simeq 1$ is the CP violating phase in the quark mass
matrix 
(a similar result would hold in the case of $\ur-\tr$ mixing, with the
CP violating source terms being proportional to
$A\lambda^3$ instead of $A\lambda^2$).
Eq.~(\ref{cpsource}) is the reason for making the particular
ansatz for supersymmetry breaking 
in Eq.~(\ref{texture}) since it allows us to derive the
baryon asymmetry from the same phase $\gamma$ that is 
in the quark mass matrix, and that is (indirectly) 
measured in the Kaon decay experiments. 
Notice that in our ansatz the quantities $x$ and 
$y$ in $M^2_{\tilde u_{LR}}$ are related to
$c$ and $s$, the mixing angles in the Unitary matrix that diagonalizes
$M^2_{\tilde u_{RR}}$ and for example, 
$x=y \Rightarrow c=s$ and the CP violating source term
vanishes [this would correspond to $A_{U_{LR}}$ being proportional to 
$A_{U_{RR}}$ in Eq. (\ref{jcp})]. In a more general case, the
parameters $x$ and $y$ present in $A_{U_{LR}}$ do not have to be
related to $c$ and $s$ which come from diagonalizing $A_{U_{RR}}$.
For example if the $stop$ is light because of a small diagonal soft
mass, and not due to mixing, we would have $c=1$, $s=0$. 

\subsection{The Rate Equations}

Once again we follow the technology explictly given in
\cite{huet-nelson} adapting their results of no family mixing to allow
for $\chr-\tr$ mixing.

The basic rate equation near equilibrium is given by 
\beq
\dot n_i = 
-\Delta_i\frac{6\Gamma_{ij}}{T^3}\Delta_j\frac{n_j}{k_j}
\eeq
where $n_i$ is a particle density,
$\Gamma$ are the reactions that change 
$n_i$ by $\Delta_i$ and $n_j$ are the
particle densities 
%with chemical potentials $\mu_j$
that participate in the reaction, changing an amount
$\Delta_j$ in the process and $k$ is a statistical factor.
Assuming super equilibrium \ie atleast some of the gauginos and
higgsinos are light enough to equilibriate quarks and squarks  
the particle densities of interest (those that participate in fast
interactions due to the large top quark Yukawa coupling) are:
\beqa
Q&=&(t_L+b_L+c_L+s_L)+
    (\tilde{t}_L+\tilde{b}_L+\tilde{c}_L+\tilde{s}_L) \nonumber \\
T&=&(t_R+c_R)+(\tilde{t}_R+\tilde{c}_R) \nonumber \\
H&=& h + \tilde{h}
\eeqa
It has been shown that strong sphaleron effects are important 
\cite{guidice}, so one has to consider the other quark species
too. However, since strong sphalerons are the only interactions they feel,
we can use the fact that strong sphalerons are rapid to constrain
them.
\beq
(u_L+d_L)=-2u_R=-2d_R=-2s_R=-2b_R\equiv -2B=(Q+T)
\label{ss1}
\eeq
in addition all the squarks of the above species are degenerate in our
model, so we have
\beq
(k_{u_L}+k_{d_L})=2k_{u_R}=2k_{d_R}=2k_{s_R}=2k_{B}
\label{ss2}
\eeq
Assuming all the squarks are heavy except the one light $\chr-\tr$
admixture gives us the following statistical factors
\beq
k_Q=12;~~~~k_T=12;~~~~k_B=3;~~~~k_H=12.
\eeq

Now we set up the rate equations for the particle species $Q$, $T$,
and $H$, including 
the effects of diffusion, the CP violating source term, and 
the ``fast'' interactions due to the top quark
Yukawa, the top quark mass, the Higgs self interaction, and the strong
sphalerons ($\Gamma_m$, $\Gamma_y$, $\Gamma_h$, and $\Gamma_{ss}$
where we redefine $6\Gamma/T^3 \rightarrow \Gamma$ for convenience).
\newpage
\beqa
\dot{Q}&=& D_q\nabla^2 Q - \Gamma_y[\frac{Q}{k_Q} - \frac{H}{k_H} -
\frac{T}{k_T}] - \Gamma_m[\frac{Q}{k_Q} - \frac{T}{k_T}] \nonumber \\
&& -6\Gamma_{ss}[4\frac{Q}{k_Q} - 2\frac{T}{k_T}+3\frac{Q+T}{k_B}]
+ \gamma_{\tilde{Q}} \nonumber \\
\dot{T}&=& D_q\nabla^2 T - \Gamma_y[-\frac{Q}{k_Q} + \frac{H}{k_H} +
\frac{T}{k_T}] - \Gamma_m[-\frac{Q}{k_Q} + \frac{T}{k_T}] \nonumber \\
&& +3\Gamma_{ss}[4\frac{Q}{k_Q} - 2\frac{T}{k_T}+3\frac{Q+T}{k_B}]
- \gamma_{\tilde{Q}} \nonumber \\
\dot{H} &=&D_h\nabla^2 H - \Gamma_y[-\frac{Q}{k_Q} + \frac{H}{k_H} 
+\frac{T}{k_T}] - \Gamma_H\frac{H}{k_H}
\label{rate}
\eeqa
Further, since the rates $\Gamma_y$ and $\Gamma_{ss}$ are independent
of the Higgs vev, these will always be fast, and we approximate the
combination of particle species feeling these interactions to be given
by their equiilibrium values \ie
\beqa
\displaystyle{\frac{Q}{k_Q} - \frac{H}{k_H} - \frac{T}{k_T}} &=& 0
\nonumber \\
\displaystyle{4\frac{Q}{k_Q} - 2\frac{T}{k_T}+3\frac{Q+T}{k_B}}&=&0
\eeqa
which gives 
\beqa
Q&=&\displaystyle{H\frac{k_Q}{k_H}\frac{3k_T-2k_B}{2k_B+3k_Q+3k_T}}
\nonumber \\
T&=&\displaystyle{-H\frac{k_T}{k_H}\frac{3k_Q+4k_B}{2k_B+3k_Q+3k_T}} 
\label{qt-h}
\eeqa
Plugging these equations into the rate equations, we can solve for the
Higgs number $H$ in the symmetric phase under the approximations
detailed in \cite{huet-nelson} to get
\beq
H=A_He^{v_w z/\bar{D}}
\label{hnum}
\eeq
where $v_W$ is the wall velocity and
\beq
A_H=\frac{\bar{\gamma} w}{\sqrt{\bar{\Gamma}\bar{D}}}
\eeq
and 
\beq
\bar{D}=\frac{D_Q(3k_Qk_T+2k_Qk_B+8k_Tk_B)
             + D_Hk_H(2k_B+3k_Q+3k_T)}
             {3k_Qk_T+2k_Qk_B+8k_Tk_B + k_H(2k_B+3k_Q+3k_T)}
\eeq
\beq
\bar{\Gamma}=(\Gamma_m+\Gamma_h)\frac{3k_Q+3k_T+2k_B}
             {3k_Qk_T+2k_Qk_B+8k_Tk_B + k_H(2k_B+3k_Q+3k_T)}
\eeq
\beq
\bar{\gamma}=\gamma_{\tilde{Q}}\frac{k_H(2k_B+3k_Q+3k_T)}
                    {3k_Qk_T+2k_Qk_B+8k_Tk_B + k_H(2k_B+3k_Q+3k_T)}
\eeq
with \cite{huet-nelson}
\beqa
D_Q \simeq {\displaystyle{\frac{6}{T}}} & 
D_H \simeq {\displaystyle{\frac{110}{T}}} \nonumber \\
\Gamma_m  \simeq 
{\displaystyle{\frac{4M_W^2\lambda_t^2\sin^2\beta}{21g^2T}}} &
\Gamma_h  \simeq  {\displaystyle{\frac{M_W^2}{35g^2T}}}
\eeqa
Using Eqs.~(\ref{ss1}) and Eqs.~(\ref{qt-h}) we can solve for
the left-handed quark density
\beq
n_L = Q + (u_L + d_L) 
    = H \displaystyle{
          \frac{3k_Qk_T-4k_Qk_B-4k_Tk_B}
               {k_H(2k_B+3k_Q+3k_T)}} 
    = \displaystyle{\frac{2}{13}H}
\eeq
Finally we turn on the 'slow' weak sphaleron rate $\Gamma_{ws}$ and
use the well known fact that an excess of left-handed baryons acts as
a chemical potential that biases the baryon number violating weak
sphalerons to produce a net baryon asymmetry. If we assume that weak
sphalerons are active in the symmetric phase, and shut off abruptly at
the bubble wall, then the solution of the rate equation for baryon
number gives
\beq
\rho_B = \displaystyle
     {-\frac{3\Gamma_{ws}}{v_w}\int_{-\infty}^{0} n_L(z)dz} 
      = \displaystyle
     {-\frac{6}{13}\frac{A_H\bar D}{v_w^2}}\Gamma_{ws}
\eeq
where we have used Eq.~(\ref{hnum}) to get the final result.
Finally, using $\lambda_t = 1,~\Gamma_{ws}=\alpha_w^4T,~
|V_{cb}|=0.04, \tan\beta=1,$ and $s=55T^3$
we get the numerical result
\beq
\frac{\rho_B}{s}=4\times 10^{-9}\frac{T}{m}\frac{e^{m/T}}{{(1-e^{m/T})}^2}
\frac{m_0\mu}{T^2}\frac{\gamma_w}{v_w}\Delta\beta
[\sin\gamma(xs^2-yc^2)]
\label{bau}
\eeq
where $m$ is the mass of the lightest 
right-handed up-type squark
in the symmetric phase, and the rest
of the parameters have been defined before. $T,~\gamma_w,~v_w$ and
$\Delta\beta$ are parameters relating to the EWPT and cannot be
measured. The rest of the parameters in Eq.~(\ref{bau}) will
presumably be measured in terrestrial experiments. 
$T$ is pretty robust around $100$ GeV. Estimates for the wall velocity
are $v_w \simeq 0.1 - 0.3$.
$\Delta\beta$ is harder to estimate: in general we would expect
some suppression depending on the ratio of physical Higgs boson
masses and we naively estimate $\Delta\beta=(m_h/m_{A})^2$, where
$m_A$ is the mass of the pseudoscalar Higgs boson.
 
\subsection{Results}

In Fig. 1 we plot the regions in the $x-y$ plane that
generate sufficient baryon asymmetry for $\chr-\tr$ or
$\ur-\tr$ mixing for two different values of $m_0$.
We use $T=100$ GeV, $\Delta\beta = 0.25$,
$v_w=0.1$, and set the CP violating phase $\gamma = \pi /4$. 
For the allowed region we require 
$\rho_B/s > 1 \times 10^{-11}$ and $85$ Gev $< m_{\tilde u_1}< 175$
GeV, where $\tilde u_1$ is the lightest up-type
squark.

\begin{figure}
\epsfig{file=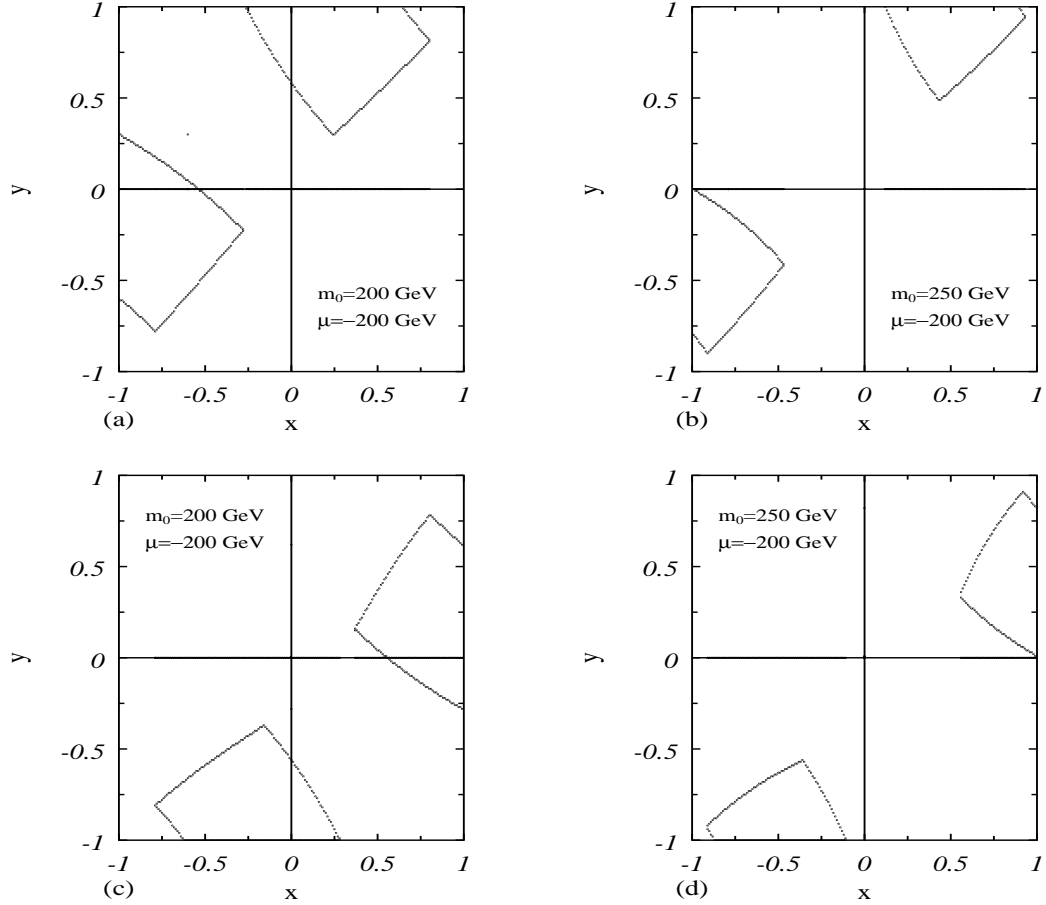,height=6in,width=6in}
\caption {Allowed regions for the mixing parameters $x$ and $y$ that
generate a sufficient baryon asymmetry. Figs. (a) and (b) are for 
$\chr - \tr$ mixing while Figs. (c) and (d) are for $\ur - \tr$
mixing.
The values of $m_0$ and $\mu$ are given in each plot, with the
rest of the MSSM parameters as in the text.}
\end{figure} 

Notice that although the magnitude of the CP violating source term is
suppressed by a factor of the Cabibbo angle for $\ur-\tr$ mixing, 
the allowed region is similar in size for the two cases. This tells us
that the dominant constraint is coming from the requirement on the
strength of the EWPT, and that the CP violation is sufficient in both
cases.
Another feature is the symmetry
under $\chr \leftrightarrow \ur$, $(x,y) \leftrightarrow -(x,y)$. 
This is because in Eq. (\ref{bau}) the dependence on $\sin\gamma$
changes sign as one goes from
$\chr-\tr$ mixing to $\ur - \tr$ mixing. 

It is clear from Fig. 1 that 
raising $m_0$ significantly decreases the allowed region. 
This is due to a combination of raising the mass of the lightest
squark above the range where the EWPT is strongly first order and the
Boltzman suppression in Eq. (\ref{bau}). 
Along these lines, if the gauginos and higgsinos are much heavier than the
phase transition temperature, $T_0 \sim 100$ GeV, we do not expect
supersymmetry to play a significant role at the EWPT. This leads us to
put an upper bound on $m_0$ and on the gaugino and higgsino 
masses of $\sim 300$ GeV, above which we do not
expect any scenario of supersymmetric baryogenesis to be viable.

The Higgs bosons affect this calculation in two ways: through the
limit placed on the lightest Higgs boson mass by the requirement of a
first order phase transition, a Higgs boson mass larger than 
$\sim 100$ GeV would
make a first order phase transition less plausible, 
and by the dependence of the baryon
asymmetry on $\Delta\beta$, the variation of the Higgs vevs over
the bubble wall. We have
estimated $\Delta\beta \sim (m_h/m_{A})^2$ and used 
$\Delta\beta = 0.25$ which results from $m_h=75$ GeV and 
$m_{A}=150$ GeV. If a more accurate
calculation shows a stronger dependence of $\Delta\beta$ on the
ratio of Higgs masses \cite {wagner2}, 
the net result would be a reduction in the baryon asymmetry.

Another possible source of suppression is the weak sphaleron rate
$\Gamma_{ws}$. We have used $\Gamma_{ws} = \alpha_w^4T^4$
\cite{sp_rate}. 
If, as suggested in some recent work
\cite{arnold}, $\Gamma_{ws}\propto \alpha_w^5T^4$, this would
significantly reduce the possibility of electroweak baryogenesis in
most models. 

Suppressing the baryon production by a factor of 10
through any of the possible mechanisms discussed above 
results in a smaller but still substantial allowed region for the
cases we have presented, except for the case of $\ur-\tr$ mixing with
$m_0=250$ GeV, where no allowed region remains (this is due to the
additional Cabibbo suppression in the CP violation). 

\section{Flavor Physics}

The largest effect in low energy flavor physics is a
large contribution to $D-\bar D$ mixing from gluino box diagrams. 
This results in an enhancement of the mixing 
by many orders of magnitude over the
\sm\ prediction. This is due to a combination of two factors. The
first is  the large supersymmetry breaking $\chr - \tr$ or $\ur-\tr$ 
mixing. The second is the fact that the Cabibbo angle is generated in
the up-type quark mass matrix [Eq. (\ref{yukawas})] coupled to the
non-degeneracy of the squark masses \cite{nir2}.
It is large over most of the allowed regions in Fig. 1, and in
fact the experimental upper bound excludes a significant amount of the
parameter space. 
Using $\Delta(m_D) < 1.3 \times 10^{-13}$ GeV,
$f_DB_D^2=160$ MeV, $m_D=1.86$ GeV and a gluino mass 
$m_{\tilde g}=150$ GeV, we plot in Fig. 3 
the allowed regions from Fig. 1 but with the added constraint
coming from $\Delta(m_D)$. 

\begin{figure}
\epsfig{file=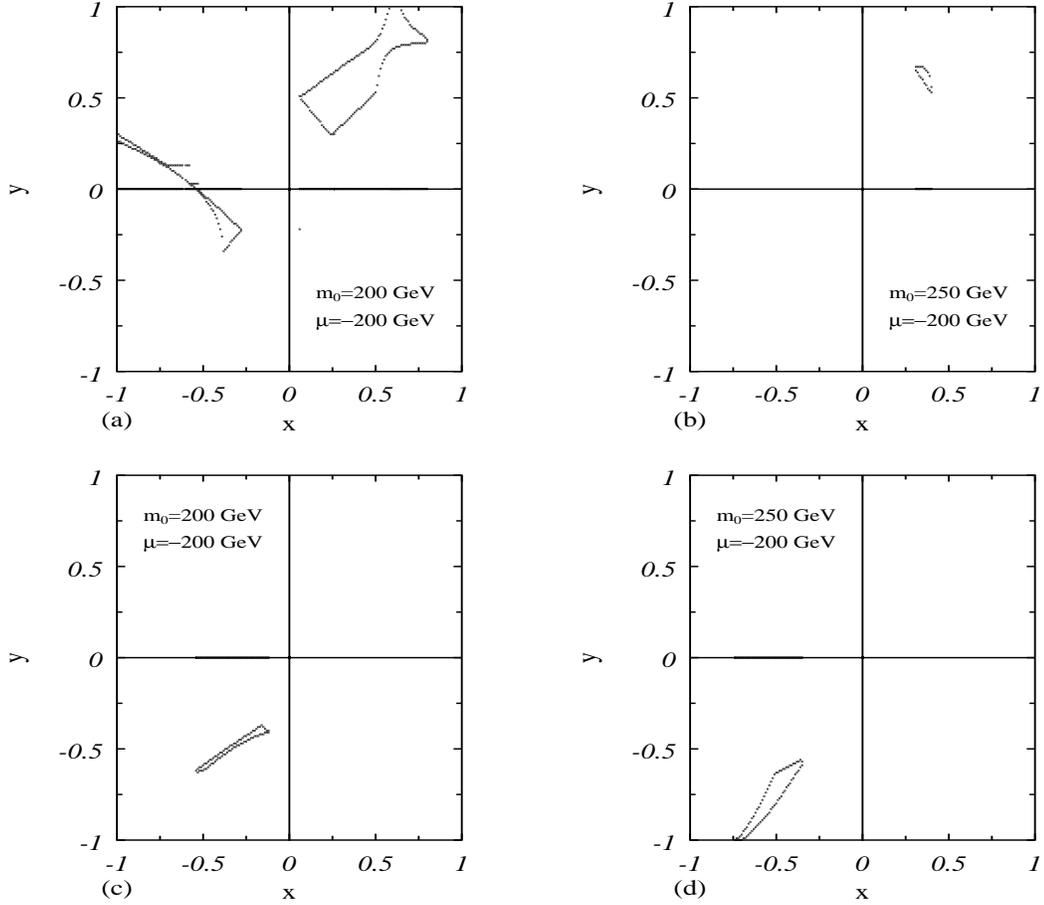,height=6in,width=6in}
\caption {Regions for the mixing parameters $x$ and $y$ that
generate a sufficient baryon asymmetry and are consistent with 
the experimental bound on $D-\bar D$ mixing. Figs. (a) and (b) are for
$\chr-\tr$ mixing while Figs. (c) and (d) are for $\ur-\tr$ mixing.
The values of $m_0$ and $\mu$ are given in each plot, with the
rest of the MSSM parameters as in the text.}
\end{figure}

Notice that a significant allowed region only remains for one case:
$\chr-\tr$ mixing and $m_0=200$ GeV. Once again, suppressing the
baryon production by a factor of 10 reduces this region slightly, but
still leaves a substantial allowed area. 
In the allowed regions, $\Delta(m_D)$ ranges
from $1 \times 10^{-14} - 1 \times 10^{-13}$ GeV, 
thus it is always within one
order of magnitude of the current upper bound, and consequently should
either be observed or rule the model out soon.

Another robust prediction of this model is a large EDM for the up
quark from one-loop gluino exchange
diagrams and for the down quark from higgsino exchange. 
Approximating the neutron EDM
by the larger of the of the up or down quark EDM's, 
we find that it ranges from $1 \times
10^{-27}$ e-cm to $1 \times 10^{-28}$ e-cm. 
once again always within a couple of
orders of magnitude of the experimental upper bound $d_n < 
1 \times 10^{-25}$ e-cm. 

The rate for the decay $b \to s \gamma$ is usually enhanced over that
in the \sm, and improvements in the experimental error, or \sm\
theoretical uncertainty would put significant constraints on the
model.

The effects on neutral $K$ and $B$ meson mixing are less pronounced,
with the dominant contribution coming from chargino boxes, with the 
result that one requires a smaller CP
violating angle $\gamma$ and $|V_{td}|$
to account for the observed CP violation in
neutral $K$ decays and the magnitude of $B_d-\bar B_d$ mixing 
respectively, changing the shape of the Unitarity triangle.
A sub leading effect is that the B-physics experiments
no longer precisely measure the angles of the Unitarity
triangle. However, this effect is pronounced only in very specific
regions of the parameter space, and even then hard to observe
experimentally \cite{worah}. 
Moreover penguin effects that could help detect new physics beyond the
\sm\ \cite{yuval-me} are negligible in this model.

If we fix the rest of the CKM matrix parameters (which don't vary much
anyway) to $\lambda=0.22$, $A=0.78$, $\sigma=0.023$, and fit for
$\gamma$ using the inputs given in Table 1 of \cite{worah}, we get 
$\gamma \simeq \pi/2$ for the \sm, and $\gamma \simeq \pi/4$ 
in our model over all of the allowed $x-y$ space. Unfortunately, given the
current hadronic uncertainties, the errors on these values for
$\gamma$ are so large that a measurement of $\gamma$ would not 
distinguish these two scenarios. In addition, because of a
conspiracy between these preferred values for $\gamma$ and $|V_{ub}|$,
the benchmark CP asymmetry measurement $B_d \to \Psi K_S$ 
will not be able to distinguish these two scenarios either
since the angle $\beta$ turns out to be similar in both cases. 
We show this in
Fig.3(a) where we plot the semi circle determined by $|V_{ub}|$ (on
which the apex of the triangle is constrained to lie), and
the Unitarity triangle corresponding to the central value for $\gamma$
in the \sm, and for a typical value of $\gamma$ in our model.
The solid parts of the semi circle correspond to the $1\sigma$ allowed
values for $\gamma$ in both models.
\begin{figure}
\epsfig{file=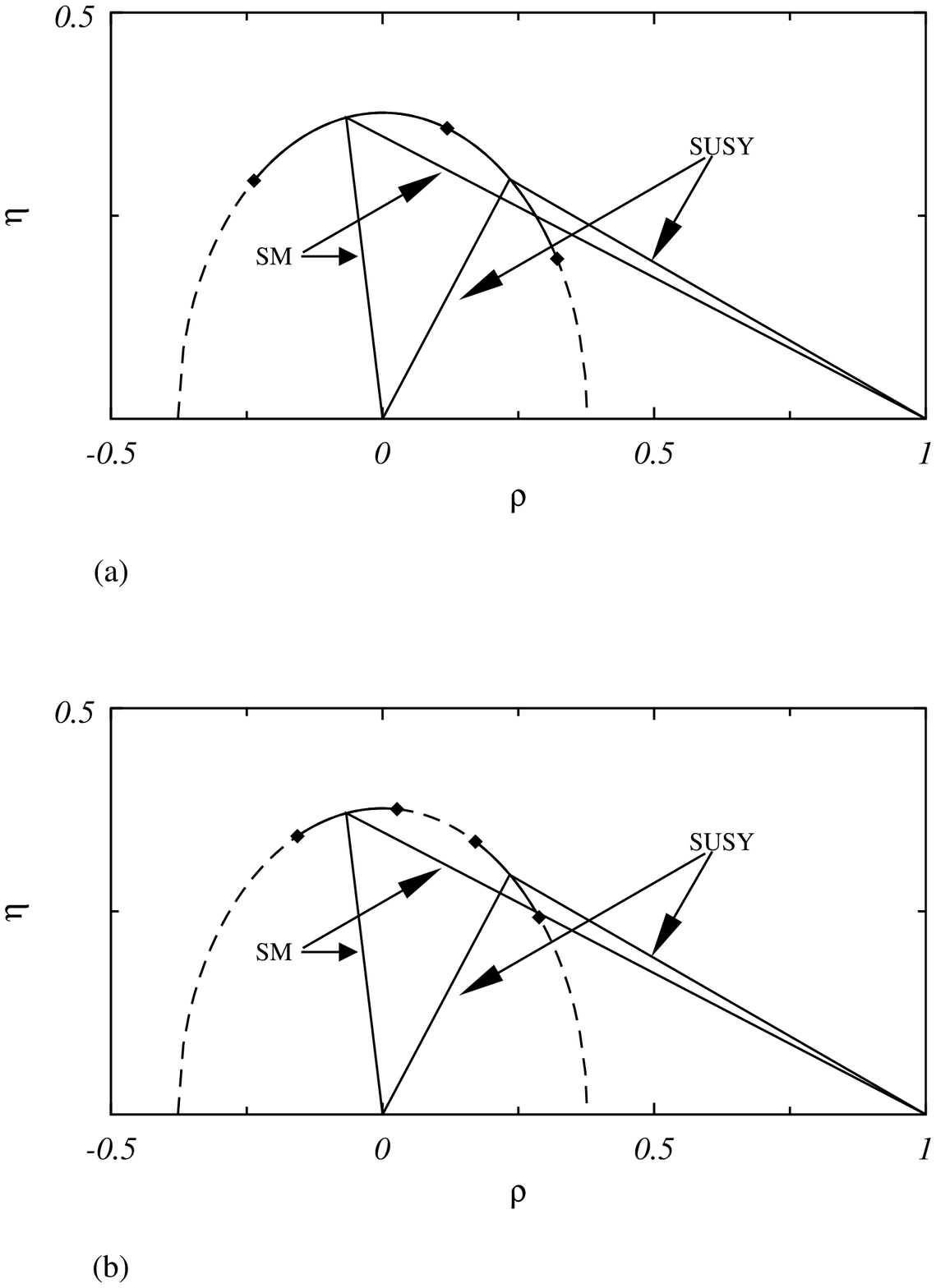,height=6in,width=6in}
\caption[a] {The Unitarity triangle in the \sm\ and in the supersymmetric
extension proposed here. The dashed semi circle is 
$|V_{ub}|$, and the apex of the triangle is constrained to lie on it.
The solid parts of the semi circle delineated by diamonds at the ends
display the $1\sigma$ 
allowed values for the angle $\gamma$ in the \sm\ and the
new model. Fig. (a) is with the current hadronic uncertainties. Fig.
(b) assumes the theoretical improvements discussed in \cite{buras}.}
\end{figure}
An improvement in calculation of hadronic matrix elements would render
the situation a little less bleak, and we plot in Fig. 3(b) the same
information as in Fig. 3(a) assuming the central values for
experimental quantities don't shift, but that the 
hadronic uncertainties are
reduced. This corresponds to Table 2 of \cite{worah}. 
In this case a direct measurement of $\gamma$, or alternatively
$\alpha=\beta+\gamma$ could differentiate the models. 
$\sin2(\beta+\gamma)$ measured in $B_d \to \pi \pi$ is clearly negative
in the \sm\ and a positive value would indicate new physics as
in our model. 

\section{Conclusions}

We have studied a particular ansatz for super symmetry breaking with
off diagonal soft masses for the up-type squarks such as generically
occur in models with Abelian flavor symmetries. The 
motivation for this is that it results in a first order EWPT,  
and makes the one
physical phase allowed in the MSSM super potential also responsible
for baryogenesis. In this scenario, the phase resides in the up-type
quark mass matrix and
has already been observed in Kaon decay experiments, and
will be well measured at the asymmetric $B$ factories. The model
predicts $D-\bar D$ mixing and a neutron EDM close to the experimental
upper bounds. The expectations for CP violating asymmetries in $B$
decays are different in this model than in the \sm, and can help
distinguish the two with an improvement in hadronic calculations.

\bigskip

{\large\bf{Acknowledgements}}

\medskip

The author wishes to acknowledge A. Grant's colloborration on parts
of this work. In addition he would like to thank
Y. Grossman, P. Huet, M. Shaposhnikov  
and S. Thomas for useful discussions.

%\newpage%
%
%{\large\bf{Note}}%
%
%\medskip%
%
%After this work was complete we were notified of a recent paper with a
%calculation of the quantity $\Delta\beta$ \cite{wagner2}. 
%Their result is an order of
%magnitude smaller than our estimate. This would not affect our results
%much as discussed in Sec. 3.4.


\begin{thebibliography}{99}

\bibitem{sakharov}
{A. D. Sakharov, {\it JETP Lett.}{\bf 5}, 24 (1967).}

\bibitem{KRS}
{V. Kuzmin, V. Rubakov and M. Shaposhnikov, \PLB{155}, 36 (1985).}

\bibitem{Farrar-Shaposhnikov}
{G. Farrar and M. Shaposhnikov, \PRL{70}, 2833 (1993); \\
\PRD{50}, 774 (1994).} 

\bibitem{Huet-Sather}
{P. Huet and E. Sather, \PRD{51}, 379 (1995).}

\bibitem{ewbaryo}
{For a review see A. Cohen, D. Kaplan and A. Nelson, \\
\ARNPS{43}, 27 (1993).}

\bibitem{const-susy}
{See for example G. Kane \etal, \PRD{49}, 6173 (1994).}

\bibitem{susy-ewpt1}
{M. Carena, M. Quiros and C. Wagner, \PLB{380}, 81 (1996);
D. Delepine \etal, \PLB{386}, 183 (1996).}

\bibitem{susy-ewpt2}
{J. Cline and K. Kainulainen, hep-ph/9605235;
M. Laine, \NPB{481}, 43 (1996).}

\bibitem{huet-nelson}
{P. Huet and A. Nelson, \PRD{53}, 4578 (1996).}

\bibitem{garristo-wells}
{R. Garisto and J. Wells, SLAC-PUB{7286}, hep-ph/9609511.}

\bibitem{luca}
{F. Gabbiani \etal, \NPB{477}, 321 (1996).}

\bibitem{nir2}
{M. Leurer, Y. Nir and N. Seiberg, \NPB{420}, 468 (1994).}

\bibitem{hor-sym}
{Y. Nir and R. Rattazzi, \PLB{382}, 363 (1996).}

\bibitem{dim-pom}
{S. Dimopoulos and A. Pomerol, \NPB{453}, 83 (1995).}

\bibitem{defMSSM}
{By Minimal Supersymmetric Standard Model we mean the supersymmetric
extension of the standard model with minimal particle content. The
constrained MSSM includes the assumptions of coupling constant
unification, universal scalar and gaugino masses and 
universal trilinear terms at a high
``unification'' scale.}

\bibitem{ckn1}
{A. Cohen, D. Kaplan and A. Nelson, \PLB{263}, 86 (1991).}

\bibitem{ckn2}
{A. Cohen, D. Kaplan and A. Nelson, \PLB{336}, 41 (1994).}

\bibitem{early-ewpt}
{S. Myint, \PLB{287}, 325 (1992); 
J. Espinosa, M. Quiros and F. Zwirner, \PLB{307}, 106 (1993).}

\bibitem{fopt}
{M.Shaposhnikov, {\it JETP Lett.}{\bf 44}, 364 (1986).}

\bibitem{espinosa}
{J. Espinosa, \NPB{475}, 273 (1996).}

\bibitem{wolf}
{L. Wolfenstein, \PRL{51}, 1945 (1983).}

\bibitem{guidice}
{G. Giudice and M. Shaposhnikov, \PLB{326}, 118 (1994).}

\bibitem{wagner2}
{M. Carena \etal, CERN-TH/96-242;hep-ph/9702409.}

\bibitem{sp_rate}
{J. Ambjorn and A. Krasnitz, \PLB{362}, 97 (1995).}

\bibitem{arnold}
{P. Huet and D. Son, UW-PT-96-20; hep-ph/9610259;
P. Arnold, D. Son and L. Yaffe, UW-PT-96-19; hep-ph/9609481.}

\bibitem{worah}
{M. Worah, \PRD{54}, 2198 (1996).}

\bibitem{yuval-me}
{Y. Grossman and M. Worah, SLAC-PUB-7351; hep-ph/9612269, to be
published in \PLB{}.}

\bibitem{buras}
{G. Buchalla, A. Buras and M. Lautenbacher, \RevModP{68}, 1125 (1996).}

\end{thebibliography}
\end{document}